\documentclass[aps,twocolumn]{revtex4} 
\usepackage{psfig} 

\newcommand\av[1]{\left<{#1}\right>}
\begin{document}

 \title{Fluctuations in network dynamics}
 
 \author{M. Argollo de Menezes and A.-L. Barab\'asi}
 \affiliation{Department of Physics, University of Notre Dame, Notre Dame,
 IN 46556}
 \date{\today}

\begin{abstract}
  Most complex networks serve as conduits for various dynamical
  processes, ranging from mass transfer by chemical reactions in the
  cell to packet transfer on the Internet. We collected data on the
  time dependent activity of five natural and technological networks,
  finding that for each the coupling of the flux fluctuations with the
  total flux on individual nodes obeys a unique scaling law. We show
  that the observed scaling can explain the competition between the
  system's internal collective dynamics and changes in the external
  environment, allowing us to predict the relevant scaling exponents.

\end{abstract}

\maketitle
 
Recent advances in uncovering the mechanisms shaping the topology of
complex networks ~\cite{networks} are overshadowed by our lack of
understanding of common organizing principles governing network
dynamics. In particular, we are far from understanding how the
collective behavior of often millions of nodes contribute to the
observable dynamical features of a given system, prompting us to
continue the search for dynamical organizing principles that are
common to a wide range of complex systems. To make advances in this
direction we need to complement the available network maps with data
on the time resolved activity of each node and link.
 
Traditional approaches to complex dynamical systems focus on the long
time behavior of at most a few dynamical variables, characterizing
either a single node or the system's average behavior.  To
simultaneously characterize the dynamics of thousands of nodes we
investigate the coupling between the average flux and fluctuations.
Our measurements indicate that in complex networks there is a
characteristic coupling between the average flux $\av{f_i}$ and
dispersion $\sigma_i$ of individual nodes (Fig. $1$).  To quantify
this observation we plot $\sigma_i$ for each node $i$ in function of
the average flux $\av{f_i}$ of the same node (Figs.
\ref{fig:sigma-f-0.5} \& \ref{fig:sigma-f-1}). We find that for five
systems for which extensive dynamical data is available the dispersion
depends on the average flux as

\begin{equation}
\sigma \sim \av{f}^{\alpha}.
\label{eq:scaling}
\end{equation}

\noindent Most intriguing, however, is the finding that the dynamical
exponent $\alpha$ is in the vicinity of two distinct values,
$\alpha=1/2$ (Fig. \ref{fig:sigma-f-0.5}) and $\alpha=1$ (Fig.
\ref{fig:sigma-f-1}), suggesting that diverse real systems can display
two distinct dynamical universality classes.

\begin{figure}[!ht]

\vskip 1cm
  {\centerline{\psfig{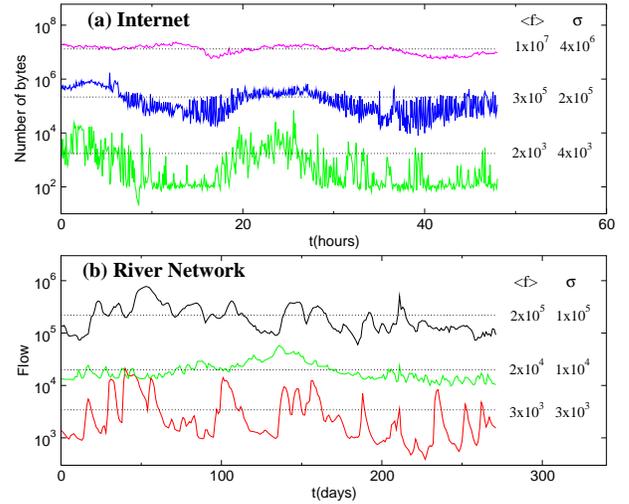}}}
  \caption{  {\bf(a)} Time dependent traffic on three
    Internet routers of the Mid-Atlantic Crossroads network, whose
    activity is monitored by the Multi Router Traffic Grapher software
    (MRTG). The figure shows the number of bytes per second for each
    of the routers in five minutes intervals for a two day period.
    {\bf(b)} Streamflow, measured in cubic feet per second, on three
    rivers in the US river basin, based on data collected by the U.S.
    Geological Survey in $2001$.  On the right of each plot we show
    the time average of the flux $\av{f}$ displayed as horizontal
    dotted lines superposed on the graphs, and the dispersion,
    $\sigma$, for each signal, indicating orders of magnitude
    differences in both flux and dispersion between nodes of the same
    network.}
  \label{fig:evolution}
\end{figure}

{\it The $\alpha\simeq 1/2$ systems (Fig. 2):} The Internet, viewed as
a network of routers linked by physical connections, serves as a
transportation network for information, carried in form of packets
~\cite{vazquez}. Daily traffic measurements of $374$ geographically
distinct routers indicate that the relationship between traffic and
dispersion follows (\ref{eq:scaling}) for close to seven orders of
magnitude with $\alpha^I = 1/2$ (Fig. \ref{fig:sigma-f-0.5}a).  In a
microprocessor, in which the connections between logic gates generate
a static network ~\cite{sole}, information is carried by electric
currents. At each clock cycle a certain subset of connections $i$ are
active, the relevant dynamical variable $f_i(t)$ taking two possible
values, $0$ or $1$. The activity during $8,862$ clock cycles on $462$
nodes of the Simple12 microprocessor indicates that the average flux
and fluctuations follow (\ref{eq:scaling}), with $\alpha^m=1/2$ (Fig.
\ref{fig:sigma-f-0.5}b).

\begin{figure}[!ht]
{\centerline{\psfig{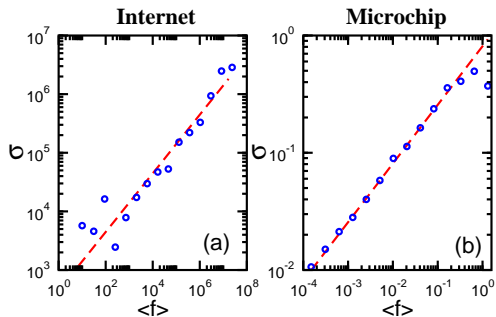}}}
\caption{The relationship between fluctuations ($\sigma$) and the
  average flux ($\av{f}$) for the $\alpha=1/2$ systems.  {\bf (a)}
  Time resolved information for $374$ Internet routers of the
  Mid-Atlantic Crossroads, ABILENE network, MIT routers, UNAM routers,
  all Brazilian RNP backbones, and dozens of smaller routers on the
  Internet, covering for each node two days of activity with five
  minute resolution.  {\bf (b)} The activity of the $462$ signal
  carriers of the 12-bit Simple12 microprocessor, recorded over
  $8,862$ clock cycles.}
  \label{fig:sigma-f-0.5}
\end{figure}

{\it The $\alpha \simeq 1$ systems (Fig. 3):} The WWW, an extensive
information depository, is a network of documents linked by URLs
~\cite{lawrence}. As many websites record individual visits, surfers
collectively contribute to a dynamical variable $f^w_i(t)$ that
represents the number of visits site $i$ receives during day $t$.  We
studied the daily breakdown of visitation for $30$ days for $3,000$
sites scattered over three continents, determining for each node $i$
the average $\av{f_i^w}$ and dispersion $\sigma_i^w$. As Fig.
\ref{fig:sigma-f-1}a shows, $\sigma_i^w$ and $\av{f_i^w}$ follow
(\ref{eq:scaling}) over three orders of magnitude with dynamical
exponent $\alpha^w = 1$. The highway system is an example of a
transportation network, the relevant dynamical variable being the
traffic at different locations. We analyzed the daily breakdown of
traffic measurements at $127$ locations on Colorado and Vermont
highways. The results, shown in Fig. \ref{fig:sigma-f-1}b, again
document scaling spanning over five orders of magnitude with
$\alpha^h=1$. Finally, the river network is a natural transportation
system ~\cite{banavar}, whose dynamics is probed via time resolved
measurements on the stream of several US rivers on $3,495$ different
locations. While these fluctuations are driven by weather patterns,
the relationship between the average stream and its fluctuations again
follows (\ref{eq:scaling}) with $\alpha^r=1$ (Fig.
\ref{fig:sigma-f-1}c).

\begin{figure}[!ht]
  {\centerline{\psfig{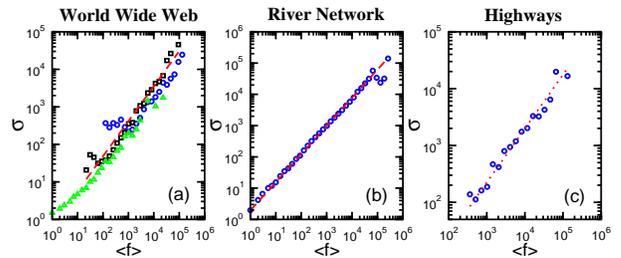}}}
  \caption{The relationship between fluctuations ($\sigma$) and the
    average flux ($\av{f}$) for systems belonging to the $\alpha=1$
    class. {\bf (a)} Daily visitations on websites collected using the
    Nedstat web monitor.  We analyzed daily traffic for a $30$ day
    period for $1,000$ sites in USA (circles) Brazil (squares) and
    Japan (triangles).  {\bf (b)} The daily streamflow of $3,945$
    rivers on the US river basin during the year of $2001$ is recorded
    by the US Geological Survey.  {\bf (c)}: Daily traffic on Colorado
    and Vermont highways representing the daily number of cars passing
    through observation points on $127$ highways from $1998$ to
    $2001$.}
  \label{fig:sigma-f-1}
\end{figure}

\begin{figure}[!ht]
{\centerline{\psfig{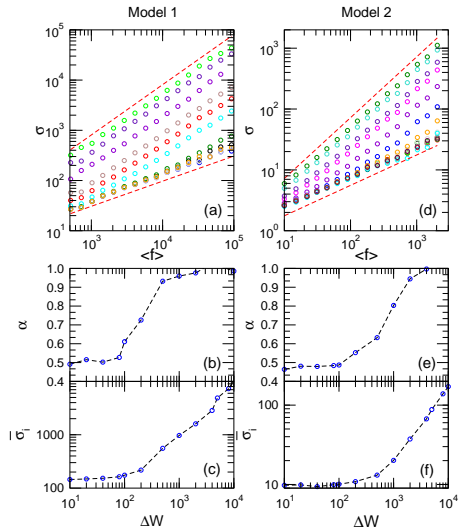}}}
\caption{In Model 1 on each ``day'' $t$ we
  release $W(t) = \av{W}+\xi (t)$ walkers on randomly selected nodes
  and allow them to perform $M=10^3$ random diffusive steps, where
  $\xi(t)$ is a uniformly distributed random variable between $-\Delta
  W$ and $\Delta W$ and $\av{W}=10^4$. {\bf (a)} The figure shows the
  $\sigma(\av{f})$ curves for $\Delta
  W=0,20,40,80,100,200,800,1000,4000,10000$ from top to bottom.  {\bf
    (b)} The dependence of the exponent $\alpha$ on $\Delta W$,
  obtained by fitting the $\sigma_i$ versus $\av{f_i}$ curves shown in
  {\bf (a)}.  Note that while the figure shows a gradual transition,
  the transition in infinite systems should be sharp.  {\bf (c)}
  Average fluctuations $\av{\sigma_i}$, obtained by averaging
  $\sigma_i$ over all nodes $i$ in the system, shown in function of
  the amplitude of the external driving force $\Delta W$.  While under
  $\Delta W\approx 10^2$ the magnitude of $\av{\sigma_i}$ is
  independent of $\Delta W$, for large $\Delta W$ the fluctuations
  increase rapidly, indicating that the network dynamics is externally
  driven. {\bf (d-f)} The same as in {\bf (a-c)}, but for Model 2,
  where the diffusive dynamics was replaced by message passing.  $W$
  was again chosen from an uniform distribution of width $\Delta W$
  and average $\av{W}=10^4$.  In all simulations we used a scale-free
  network ~\cite{ba99} with $\gamma=3$ and $10^4$ nodes.}
\label{fig:avg-sigma}
\end{figure}

To understand the origin of the observed dynamical scaling law
(\ref{eq:scaling}) we study a simple dynamical model that incorporates
some key elements of the studied systems. While the topology of these
systems vary widely, from a tree (rivers) to a scale-free network
(WWW, Internet), a common feature of the studied systems is the
existence of a transportation network that channels the flux toward
selected nodes. Therefore, we start with a network of $N$ nodes and
$L$ links, described by an adjacency matrix $M_{ij}$, which we choose
to describe either a scale-free or a random network ~\cite{networks}.
As the dynamics of the studied systems varies widely, we study two
different dynamical rules. {\it Model 1} considers the random
diffusion of $W$ walkers on the network, such that each walker that
reaches a node $i$ departs in the next time step along one of the
links the node has. Originally each walker is placed on the network at
a randomly chosen location and removed after it performs $M$ steps,
mimicking in a highly simplified fashion a human browser surfing the
Web for information. To probe the collective transport dynamics
counters attached to each node record the number of visits by various
walkers. To capture the day to day fluctuations on individual nodes we
repeat independently $D$ times the diffusion of $W$ walkers on the
same fixed network and denote by $f_i(t)$ the number of visits to node
$i$ on day $t=1,\dots,D$. As Fig.  \ref{fig:avg-sigma}a indicates, the
average flux and fluctuations follow (\ref{eq:scaling}) with
$\alpha=1/2$.  In {\it Model 2} we replaced the diffusive dynamics
with a directed flow process.  In this case each day $t$ we pick $W$
randomly selected pairs of nodes, designating one node as a sender and
the other as a recipient, and send a message between them along the
shortest path.  Counters placed on every node count the number of
messages passing through.  This dynamics mimics, in a highly schematic
fashion, the low density traffic between two nodes on the Internet.
As Fig. 4d shows, we find that Model 2 also predicts $\alpha=1/2$,
indicating that the $\alpha=1/2$ exponent is not a particular property
of the random diffusion model, but it is shared by several dynamical
rules.

We can understand the origin of the $\alpha=1/2$ exponent if we
inspect the nature of fluctuations in Model 1. In the $M=1$ limit
walkers arrive to randomly selected nodes but fail to diffuse further,
reducing the dynamics to random deposition, a well known model of
surface roughening ~\cite{barabasi-book}. Therefore, the average
visitation on each node grows linearly with time, $\av{f} \sim t$, and
the dispersion increases as $\sigma \sim t^{1/2}$, providing
$\alpha=1/2$ ~\cite{barabasi-book}. While for $M>1$ diffusion
generates correlations between the nodes, we find that the
fluctuations on the individual nodes, $\sigma_i^{int}$, continue to be
dominated by the internal randomness of the walker arrival and
diffusion process, following the $\alpha=1/2$ dynamical exponent
\cite{poisson}.

To understand the origin of the second ($\alpha=1$) universality
class, we note that in real systems the fluctuations on a given node
are determined not only by the system's internal dynamics, but also by
changes in the external environment. To incorporate externally induced
fluctuations we allow $W$ (the number of walkers and messages in
Models 1 and 2), to vary from one day to the other.  Assuming that the
day to day variations of $W(t)$ define a dynamic variable chosen from
an uniform distribution in the interval $[W-\Delta W,W+\Delta W]$, for
$\Delta W=0$ we recover $\alpha=1/2$.  However, when $\Delta W$
exceeds a certain threshold, in both models the dynamical exponent
changes to $\alpha=1$ (Fig.  \ref{fig:avg-sigma}b and e).

To understand the origin of the $\alpha=1$ exponent we notice that on
each node the observed day to day fluctuations have two sources. For
$\Delta W=0$ we have only internal fluctuations, coming from the fact
that under random diffusion (or random selection of senders and
receivers in Model 2) the number of walkers (messages) that visit a
certain node displays day to day fluctuations. For $\Delta W\neq 0$
the fluctuations have an external component as well, as when the total
number of walkers (messages) change from one day to the other, they
proportionally alter the visitation of the individual nodes as well.
If the magnitude of the day to day fluctuations is significant, they
can overshadow the internal fluctuations $\sigma_i^{int}$.  Indeed, if
in a given time frame the total number of walkers or messages doubles,
the flux on {\it each node} is expected to grow proportionally, a
potentially much larger variation than the changes induced by the
internal fluctuations. Therefore, for $\Delta W \ne 0$ the external
driving force, determined by the time dependent $W(t)$, contributes to
the daily fluctuations with a dispersion $\sigma^{dr}(\Delta W) =
\sqrt{\av{{W(t)^2}} - \av{W(t)}^2}$. The total fluctuations for node
$i$ are therefore given by $\sigma_i ^2=\left(\sigma_i^{int}\right)^2+
\left(\sigma_i^{ext}\right)^2$. As the effect of the driving force is
felt to a different degree on each node, we can write $\sigma_i^{ext}=
A_i\sigma^{dr}(\Delta W)$, where $A_i$ is a geometric factor capturing
the fraction of walkers channeled to node $i$, and depends only on the
position of node $i$ within the network.  When $\Delta W=0$, the
external component $\sigma^{dr}$ vanishes, resulting in
$\sigma^{int}_i=a_i \av{f_i}^{1/2}$, as discussed earlier, where $a_i$
is an empirically determined coefficient. When $\Delta W$ is
sufficiently large, so that $A_i\sigma^{dr}(\Delta W) \gg
\sigma_i^{int}$, then the fluctuations on each node are dominated by
the changes in the external driving force. In this limit a node's
dynamical activity mimics the changes in the external driving force,
allowing us to approximate the flux at node $i$ with $f_i(t) = A_i
W(t)$. In this case we have $\av{f_i} = A_i \av{W(t)}$ and
$\av{{f_i}^2} = {A_i}^2 \av{{W(t)}^2}$, giving $\sigma_i =
\sqrt{\av{{f_i}^2} - \av{f_i}^2}~=A_i\sigma^{dr}$.  As $\sigma^{dr}$
and $\av{W(t)}$ are time independent characteristics of the external
driving force, we find $\sigma_i \simeq \sigma_i^{ext} =
\frac{\sigma^{dr}}{\av{W(t)}} \av{f_i}$, providing the observed
coupling (\ref{eq:scaling}) with $\alpha=1$. Note that this derivation
is independent of the network topology or the transport process,
predicting that any system for which the magnitude of fluctuations in
the external driving force exceeds the internal fluctuations will be
characterized by an $\alpha=1$ exponent.

These calculations imply that the fluctuations on a given node can be
decomposed into an internal and an external component as
\begin{equation}
\sigma_i ^2 = a_i^2 \av{f_i} +
\left(\frac{\sigma^{dr}}{\av{W(t)}}\av{f_i}\right)^2.
\label{eq:sigma_i}
\end{equation}

\noindent Therefore, increasing the amplitude of
fluctuations $\Delta W$ should induce a change from the $\alpha=1/2$
intrinsic or endogenous to the $\alpha=1$ driven behavior. To confirm
the validity of this prediction, in Figs.  4c and f we show the
average fluctuation $\bar{\sigma_i}$ over all nodes in function of the
amplitude $\Delta W$ of the driving force.  For both models we find
that for small $\Delta W$ values $\bar{\sigma_i}$ remains unchanged,
as in this regime $\bar{\sigma_i} \sim \sigma^{int}_i >
\sigma^{ext}_i$.  However, after $\Delta W$ exceeds a certain
threshold, $\bar{\sigma_i}$ changes behavior, monotonically increasing
with $\Delta W$. In this second regime the fluctuations are driven by
external forces, $\bar{\sigma_i}\sim \sigma^{ext}_i \sim
\bar{A_i}\sigma^{dr}$, and according to (\ref{eq:sigma_i}) we should
observe $\alpha=1$. Indeed, we find that in both models the transition
from the constant to the increasing $\bar{\sigma_i}$ regime (Figs.
4c,f) coincides with the crossover from the $\alpha=1/2$ to $\alpha=1$
(Figs. 4b,e). Note, however, that the gradual transition observed in
Figs.  4b-e from $\alpha=1/2$ to $\alpha=1$ is a numerical artifact of
the fitting process: in the transition regime the $\alpha=1/2$ and
$\alpha=1$ scaling coexist on the same $\sigma(\left<f\right>)$ curve,
giving an exponent that is different from $1/2$ or $1$. In reality the
transition between the two regimes is sharp.  To understand to what
degree our findings depend on the specific simulation and model
details we changed the topology from scale-free \cite{ba99} to random
network and from undirected to directed network, as well as altering
the nature of the external fluctuations by keeping $W$ constant in
Model 1 but forcing the number of steps, $M$, to play the role of the
stochastic external driving force. For each version we recover the
transition between the $\alpha=1/2$ and $\alpha=1$ when the amplitude
of the external fluctuations exceeds a certain threshold
\cite{comment}.

These results indicate that the $\alpha=1/2$ exponent captures an
endogenous behavior, determined by the system's internal collective
fluctuations. In the studied model internal fluctuations are rooted in
the randomness in the walkers' arrival and diffusion; on the Internet
they originate in the choices users make to where and when to send a
message; for the computer chip they come from the alternating
utilization of the various circuits, as required by the performed
computation.  In contrast, the $\alpha=1$ exponent describes driven
systems, in which the fluctuations of individual nodes are dominated
by time dependent changes in the external driving forces. Therefore,
fluctuations of World Wide Web traffic, river streams and highway
traffic are driven by such external factors as daily variations in the
number of Web surfers, seasonal or daily changes in precipitation or
daily variations in the number of drivers, respectively.

Of the two observed exponents our derivation indicates that $\alpha=1$
is universal, being independent of the nature of the internal dynamics
or the network topology. There are no firm restrictions, however, on
the scaling of the internal dynamics, raising the possibility that
self-organized processes could lead to collective fluctuations that
are characterized by $\alpha$ exponents different from $1/2$.
Empirical evidence for potential intermediate $\alpha$ values comes
from ecology, where (\ref{eq:scaling}) describes spatial and temporal
variations of populations ~\cite{taylor}. It is much debated, however,
whether the observed scaling represent valid exponents, or only
crossovers between $\alpha=1/2$ and $1$ ~\cite{anderson}.

We are indebted to Jay Brockman and Steven Balensiefer for providing
the data on the computer chip.  This research was supported by grants
from NSF, NIH and DOE.


\begin{thebibliography}{99}
\bibitem{poisson} While Poissonian statistics explains the
  $\alpha=1/2$ scaling in the random deposition model, it is not
  appropriate for Ethernet traffic \cite{taqqu}, indicating that the
  Internet's $\alpha=1/2$ exponent is rooted in the interplay between
  network topology and dynamics. The exponential distribution
  describing waiting times is known to predict $\alpha=1$.
\bibitem{comment} The relative strength of fluctuations can be
  determined independently of the scaling exponents. Such measurements
  indicate that for the Internet and the microchip the magnitude of
  the internal fluctuations is close to $100$ times larger than the
  magnitude of the external fluctuations.  In contrast, for the other
  systems the internal and external fluctuations are comparable in
  magnitude [M.  Argollo de Menezes and A.-L. Barab\'asi, to be
  published].
\end{thebibliography}
\end{document}